\documentclass[%
 reprint,
superscriptaddress,
 amsmath,amssymb,
 aps,
]{revtex4-2}
\usepackage{physics}
\usepackage{graphicx}
\usepackage{dcolumn}
\usepackage{bm}
\usepackage[colorlinks, citecolor=blue]{hyperref}

\begin{document}

\title{Tomographic markers and photon addition to coherent states of light: Comparison with experiment}

\author{Soumyabrata Paul}
\email{soumyabrata@physics.iitm.ac.in}
\affiliation{Department of Physics, Indian Institute of Technology Madras, Chennai 600036, India}
\affiliation{Center for Quantum Information, Communication and Computing (CQuICC), Indian Institute of Technology Madras, Chennai 600036, India}
\author{S. Lakshmibala}
\affiliation{Center for Quantum Information, Communication and Computing (CQuICC), Indian Institute of Technology Madras, Chennai 600036, India}
\author{V. Balakrishnan}
\affiliation{Center for Quantum Information, Communication and Computing (CQuICC), Indian Institute of Technology Madras, Chennai 600036, India}
\author{S. Ramanan}
\affiliation{Department of Physics, Indian Institute of Technology Madras, Chennai 600036, India}
\affiliation{Center for Quantum Information, Communication and Computing (CQuICC), Indian Institute of Technology Madras, Chennai 600036, India}

\date{\today}

\begin{abstract}
Photon addition to quantized light is of immense interest, both experimentally and theoretically. We identify a set of markers that play an important role in the context of photon addition to coherent states of light. These markers are directly computable from optical tomograms. We calculate the amplification gain due to photon addition, and the dependence of quadrature variances on relevant parameters, from the tomograms and compare them with results obtained after state reconstruction in a recent experiment. Our results match well with the fidelity plots reported by the experimenters. Our approach which circumvents state reconstruction could provide a viable procedure to identify specific aspects of photon addition to nonclassical light as well, from the tomograms themselves.
\end{abstract}

\maketitle


\section{Introduction\label{sec:introduction}}
Photon addition to various quantum states of the radiation field has both theoretical and experimental implications. Photon added states are potentially good candidates for enhancing nonclassical effects, quantum sensing and metrology. The popular single mode field states for photon addition are the standard coherent state of light (CS), the squeezed vacuum and the cat states.
The single photon added coherent state (1-PACS) was first proposed and studied in~\cite{Agarwal:1991}, and experimentally identified through state tomography in~\cite{Zavatta:2004, Zavatta:2005}. Since then there has been much progress in the addition of photons to different states of light, using a variety of procedures. This has culminated in successful photon by photon engineering (see, e.g.,~\cite{Biagi:2022a}), experimental realization of conditional addition of photons to the CS~\cite{Fadrny:2024},  diverse applications of non-Gaussian  light~\cite{Lvovsky:2020}, and enhancement of noise sensitivity by judicious addition of photons to the compass kitten state~\cite{Akhtar:2024}. The nonclassical properties that arise on photon addition to the squeezed CS ~\cite{Tinh:2023}, and the changes in the sub-Planck structure of the Wigner functions of cat states ~\cite{Arman:2021}, have been examined.  Several theoretical and experimental investigations on photon addition to the CS  have been reported in the literature (see, e.g.,~\cite{Marek:2008, Barbieri:2010, Zavatta:2011, Kumar:2013, Biagi:2020, Biagi:2021, Biagi:2022b}). All of these are based on the density operator/Wigner function which is reconstructed by quantum state tomography (QST) from tomograms. However, QST is difficult in most continuous variable systems, because of the large dimension of the Hilbert space. In general, computation of the off-diagonal elements of the density operator from the tomogram poses challenges. It is therefore of interest to identify a viable procedure to detect the extent of photon addition to a state, with markers that can be deduced directly from the tomogram. The tomogram is a quorum of probability density functions (PDFs) (equivalently, the diagonal elements of the density operator in the basis of the field quadrature $X_{\theta}$) obtained at different angles $\theta$  in a homodyne set up. In principle, all information about the quantum state can be extracted even if only the diagonal elements of the density operator 
are measured for every value of $\theta$ (i.e., corresponding to an infinite number of tomographic slices)~\cite{Ibort:2009}. In practice however, depending on the properties of the relevant state, a finite number of judiciously chosen, experimentally accessible  slices suffices. This has been  demonstrated earlier through procedures to estimate nonclassical effects such as the extent of squeezing~\cite{Wunsche:1996}, and to compute entanglement indicators~\cite{Sharmila:2017}, directly from tomograms.

It has been established theoretically~\cite{Soumyabrata:2025} that photon addition to the single mode squeezed vacuum and to the cat states is manifested qualitatively through  changes in their  optical tomograms. Further, the  Wasserstein distance  $W_{1}$~\cite{Vaserstein:1969} is a useful tomographic marker in this context. However, $W_{1}$ is merely one of many quantifiers of the distance between two PDFs. It is  important to assess the efficacy of other similar quantifiers as well, not merely theoretically but also by comparison with experimental findings. In what follows we analyse photon addition to the CS, and compare the performance of $W_{1}$ with that of the Kullback-Leibler  divergence ($D_{\rm KL}$)~\cite{KullbackLeibler:1951} and the Bhattacharyya distance ($D_{\rm B}$)~\cite{Bhattacharyya:1943}, in identifying the extent of difference in the PDFs of the CS and its photon added counterparts. We corroborate our results with 
a recent experimental report~\cite{Fadrny:2024}.

The plan of the paper is as follows. In Sec.~\ref{sec:tomogram_WD_DKL_BD} we introduce the salient features of optical tomograms, define $W_{1}$, $D_{\rm KL}$ and $D_{\rm B}$ and point out their use in comparing tomographic patterns. In Sec.~\ref{sec:Fadrny_PACS_expt} we focus on photon addition to the CS. We compute these three quantifiers directly from appropriate tomograms, treating the CS as the reference state, and assess their relative merits by comparison with recently reported experimental results~\cite{Fadrny:2024}. We conclude with a brief summary and outlook in Sec.~\ref{sec:Summary}. Details of the calculations are elaborated upon in the Supplementary Material (SM).

\section{Optical tomograms and distances between probability distributions}\label{sec:tomogram_WD_DKL_BD}
Given a single-mode radiation field with photon creation and annihilation operators ($\hat{a}^{\dagger}, \hat{a}$), we consider the set of rotated quadrature operators~\cite{Ibort:2009}
$\hat{\mathbb{X}}_{\theta} = \big( \hat{a}^{\dagger} e^{i\theta} + \hat{a} e^{-i\theta} \big) / \sqrt{2}$,
where $\theta \in$ [$0, \pi$) is the phase of the local oscillator in the standard homodyne measurement setup. $\theta = 0$ and $\pi/2$ respectively correspond to the $x$ and $p$ quadratures. The set $\{{\hat{\mathbb{X}}_{\theta}\}}$ constitutes a quorum of observables carrying complete information about a given state with density matrix $\hat{\rho}$. The optical tomogram $w(X_{\theta}, \theta)$ is given by \cite{Lvovsky:2009}
\begin{equation} 
w(X_{\theta}, \theta) = \langle X_{\theta}, \theta |\, \hat{\rho} \,| \ X_{\theta}, \theta \rangle\;\text{where}\;
\hat{\mathbb{X}}_{\theta} |X_{\theta}, \theta \rangle = X_{\theta}|X_{\theta}, \theta \rangle.
\label{eq:tase_w_single}
\end{equation}
Here \{$|X_{\theta}, \theta \rangle$\} forms a complete basis (a continuous quadrature basis) for a given $\theta$. The quadrature tomogram (which is the object with which we will be primarily concerned) is thus a collection of histograms corresponding to the quadrature operators. It satisfies the relations
$\int_{-\infty}^{\infty}\, dX_{\theta}\, w(X_{\theta}, \theta) = 1$
\label{eq:tase_w_single_norm}
for every $\theta$ (completeness) and $w(X_{\theta}, \theta + \pi) = w(-X_{\theta}, \theta)$ (symmetry). For a pure state $\hat{\rho} = |\psi\rangle \langle\psi|$, and  for $\theta  = 0$ (say),~\eqref{eq:tase_w_single} simplifies to $w(X_{0}, 0) = |\psi(x)|^{2}$, the PDF in the $x$ quadrature. For a general quadrature specified by $\theta$, $\hat{\rho} = |\psi(X_{\theta}, \theta)|^{2}$. Note that the tomogram comprises {\it only} the diagonal elements of $\hat{\rho}$ in any basis. The optical tomogram is presented with $X_{\theta}$ as the abscissa and $\theta$ as the ordinate (see Figs.~1,~4 and~9 of SM). Thus, a single-mode optical tomogram is essentially a collection of one-dimensional probability distributions, each corresponding to a different value of $\theta$, assembled together to form a \textit{pattern}. Pattern comparisons are readily carried out with the Wasserstein distance as a quantifier, as explained below.

Consider two normalized PDFs $f(x)$ and $g(x)$, with  corresponding cumulative distribution functions (CDFs) $F(x)=\int_{-\infty}^{x}~dy~f(y)$ and $G(x)=\int_{-\infty}^{x}~dy~g(y)$. The Wasserstein distance (more accurately, the $1$-Wasserstein distance) between $F(x)$ and $G(x)$ is given by 
\begin{equation}
W_{1}(F, G) = \int_{-\infty}^{\infty} \!\!\! dx \left| F(x) - G(x) \right|.
\label{eq:def_wassdist_1}
\end{equation}

\noindent
The computation of $W_{1}$  between the Husimi distributions (obtained after state reconstruction) corresponding to several standard states of light has been given in~\cite{Zyczkowski:1998}. Our approach, on the other hand, involves a direct computation of the distances from the tomograms themselves. The advantage of this approach has already been stated.

Apart from $W_{1}$, both $D_{\rm KL}$ and $D_{\rm B}$ are often used to quantify the difference between two probability distributions. They are defined as
\begin{equation}
D_{\rm KL}(f,g) = \int_{-\infty}^{\infty}\!\!\! dx\, f(x)\, \mathrm{ln}\, [f(x)/g(x)],
\label{eq:def_KL}
\end{equation}
\begin{equation}
D_{\rm B}(f,g) = -\mathrm{ln} \int_{-\infty}^{\infty}\!\!\! dx \, [f(x) g(x)]^{1/2}.
\label{eq:def_BD}
\end{equation}

\section{Discriminating between different photon added coherent states from optical tomograms: Comparison with experiment\label{sec:Fadrny_PACS_expt}}
\begin{figure}
\centering
\includegraphics[width=0.37\textwidth]{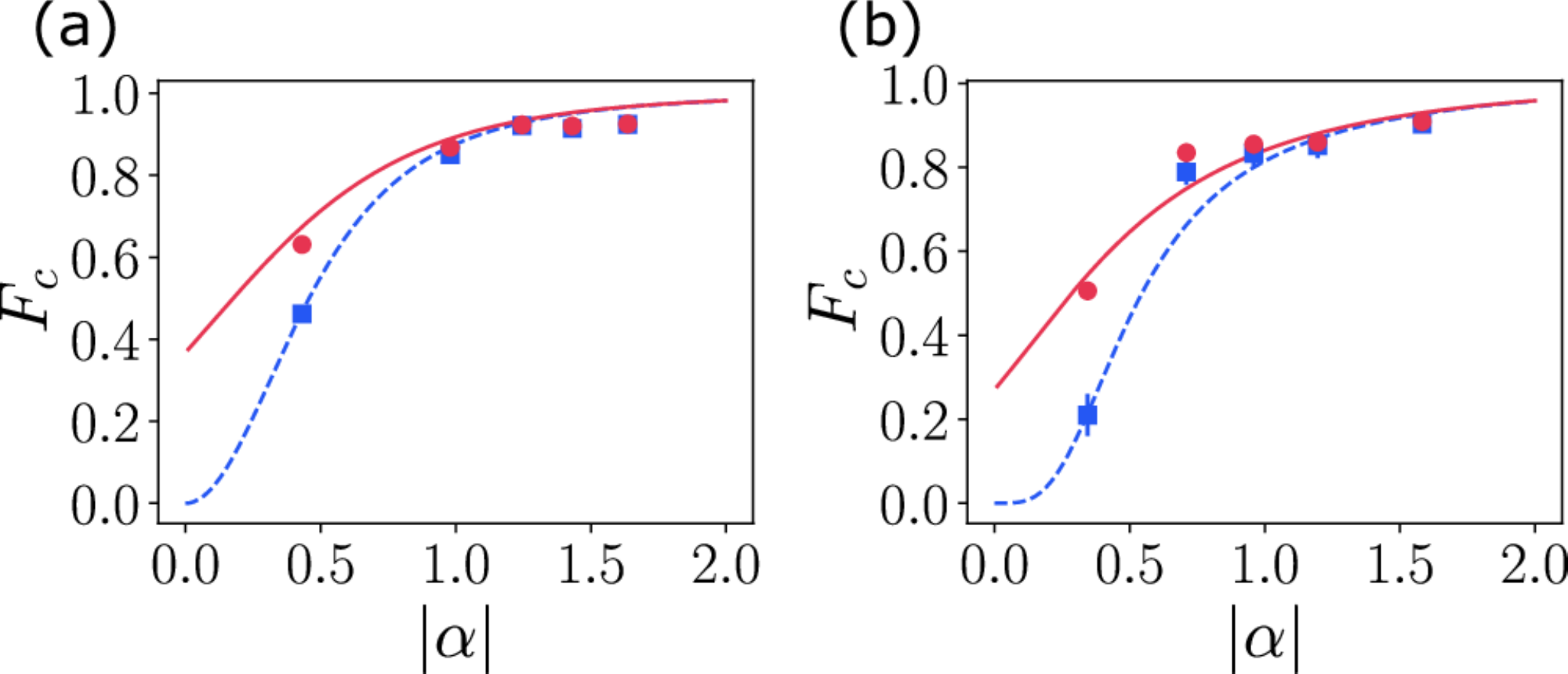}
\caption{{\bf Fidelities of $n$-photon-added coherent states with exact coherent states}. Red solid lines represent the theoretical dependence of fidelity with the coherent state with amplitude $\beta_{\rm opt}$ given by Eq.~(11) of~\cite{Fadrny:2024}. Blue dashed lines depict theoretical fidelities with the coherent states with amplitude ${\sf g}_{n}(\alpha)\alpha$, where ${\sf g}_{n}(\alpha)$ is the amplification gain (see Eq.~(7) of~\cite{Fadrny:2024}). Red circles and blue squares are the corresponding experimental fidelities determined from the reconstructed density matrices of the generated states. Data are plotted for $n = 1$ ({\bf a}) and $n = 2$ ({\bf b}). \\
\vspace{1ex}
This figure (with caption) has been taken from the publication ``Experimental preparation of multiphoton-added coherent states of light" by J. Fadrn{\'y}, M. Neset, M. Bielak, M. Je{\v{z}}ek, J. B{\'i}lek, and J. Fiur{\'a}{\v{s}}ek, npj Quantum Information {\bf 10}, 89 (2024)~\cite{Fadrny:2024}, DOI:10.1038/s41534-024-00885-y.}
\label{fig:Fadrny2024_figure_5}
\end{figure}

The normalized $m$-photon added CS ($m$-PACS) $|\alpha,m\rangle$ obtained from the CS  $|\alpha\rangle $ is given by 
$|\alpha, m\rangle = a^{\dagger m}|\alpha\rangle / [ m!L_{m}(-|\alpha|^{2})]^{1/2}$,
\label{eq:pacs_expression}
where $L_{m}$ is the Laguerre polynomial of order $m$. In this section we compare the results obtained solely from tomograms, in the context of photon addition to the CS $|\alpha\rangle$, with recent experimental results~\cite{Fadrny:2024}. In this experiment the authors have demonstrated the implementation of conditional addition of up to $3$ photons to $|\alpha\rangle$, reconstructing the states/Wigner functions using time-domain homodyne tomography. Of relevance to us are Figs.~4 and~5 of~\cite{Fadrny:2024}. For ready reference these figures have been reproduced in Fig.~6 of SM and Fig.~\ref{fig:Fadrny2024_figure_5} here. We will be concerned with three aspects, numbered  (i), (ii) and (iii), discussed below. \\

\noindent
(i) {\it Plots of the amplification gain ${\sf g}_{m}(|\alpha|)=\langle \alpha,m|a|\alpha,m\rangle/|\alpha|$ due to addition of photons, and quadrature variances, as functions of $|\alpha|$}:
For ease of comparison between our results and those from the experiment (Figs.~4(a)--(d) of~\cite{Fadrny:2024}) on amplification gain and quadrature variances, we have presented all the relevant plots in Figs.~5--7 of SM. We summarize the results here.

In the experiment, ${\sf g}_{m}(|\alpha|)$ has been explicitly computed from the reconstructed states $|\alpha,m\rangle$ as functions of $|\alpha|$ (see Figs.~7(a) and~(b) of SM). It has been shown that in both the theoretical computation using the explicit expression for $|\alpha,m\rangle$ and the experimental data for the range of values of $|\alpha|$ considered ($0.5 \leqslant |\alpha| \leqslant 2$), ${\sf g}_{m}(|\alpha|)$ is a decreasing function of $|\alpha|$. In Figs.~7(c) and~(d) of SM, the experimental plots of variances in canonically conjugate quadratures as functions of $|\alpha|$ are shown. (These are Figs.~4(c) and~(d) of~\cite{Fadrny:2024}).

Alternatively, we have computed ${\sf g}_{m}(|\alpha|)$ directly from the tomograms corresponding to $|\alpha,m\rangle$, using the result~\cite{Wunsche:1996} 
\begin{align}
\langle \hat{a}^{\dagger k}\hat{a}^{l} \rangle 
&= C_{kl} \sum_{m=0}^{k+l} \exp \Big\{ - \frac{im(k-l)\pi}{k+l+1} \Big\} \times\nonumber \\
& \int_{-\infty}^{\infty} dX_{\theta}~w \Big( X_{\theta}, \frac{m\pi}{k+l+1} \Big) H_{k+l}(X_{\theta}),
\end{align}
where
$C_{kl}=k!l!/\big\{ (k+l+1)! 2^{(k+l)/2} \big\}$ and $H_{k+l}(X_{\theta})$ is the Hermite polynomial of degree $k+l$ computed as a function of $X_{\theta}$ for a given $\theta$. Since $\theta = (m\pi)/(k+l+1)$ in the above expression, it is clear that $(k+l+1)$ slices of the tomogram including $\theta=0$ are needed to compute $\langle \hat{a}^{\dagger k}\hat{a}^{l} \rangle$. Setting $k=0$ and $l=1$, we compute 
${\sf g}_{m}(|\alpha|)$ numerically from the tomograms for the CS and the $m$-PACS ($m=1, 2$).
Representative tomograms for these states are given in Fig.~4 of SM.

Based on the foregoing discussion, relevant plots (both from the experiment, and from our calculations using the tomograms are presented in SM for ready comparison with each other). Our plots for ${\sf g}_{m}(|\alpha|)$ versus $|\alpha|$ are presented in Fig.~5 of SM. It is clear that these plots reproduce the theoretical plots obtained by explicitly using the states $|\alpha,m\rangle$ (as expected), and therefore capture the trends seen in the experimental data~\cite{Fadrny:2024} (Figs.~7(a) and~(b) of SM), to the same extent. We have also computed the quadrature variances as functions of $|\alpha|$ (Figs.~6(a) and~(b) of SM) directly from the tomograms. They agree well with the variances computed after state reconstruction: compare with Figs. 4(c) and~(d) in~\cite{Fadrny:2024} (reproduced in Figs.~7(c) and (d) of SM). \\

\begin{figure}
\centering
\includegraphics[width=0.45\textwidth]{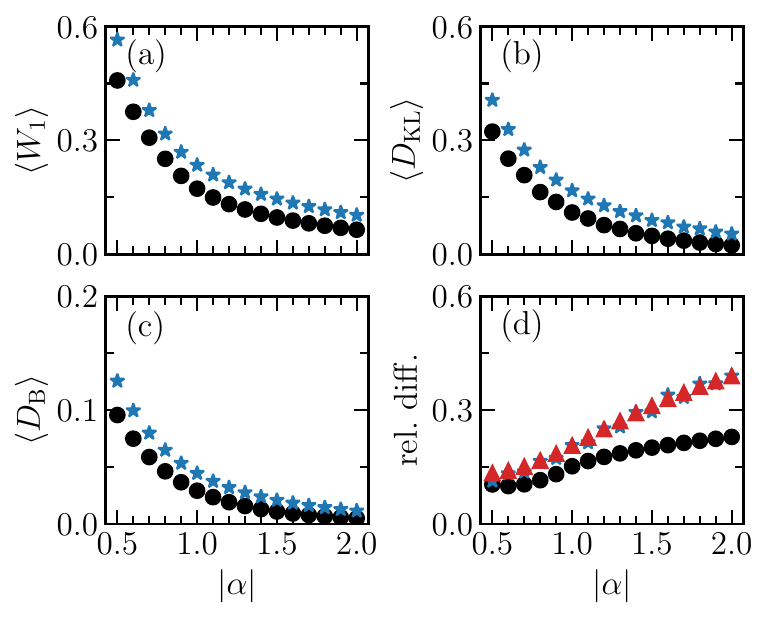}
\caption{(a) $\langle W_{1} \rangle$, (b) $\langle D_{\rm KL} \rangle$ and (c) $\langle D_{\rm B} \rangle$ are the average values of the Wasserstein distance, Kullback-Leilber divergence, and Bhattacharyya distance respectively as functions of $|\alpha|$. The averages are obtained by computing these distances between $|\alpha,m\rangle$ and $|\beta_{\rm opt}\rangle$ ($m=1$: black circles and $m=2$: blue asterisks) for $5$ different tomographic slices (i.e., $5$ different values of $\theta$ which are equally spaced between $0$ and $\pi$). Increasing the number of tomographic slices does not change the values of these three averages significantly. The three averages increase with increase in $m$ (from $1$ to $2$) for any given $|\alpha|$. With increase in $|\alpha|$, the three averages decrease, corresponding to an increase in the fidelity between $|\alpha,m\rangle$ and $|\beta_{\rm opt}\rangle$, which is borne out by the experimental plots (Figs.~5(a) and~(b) in~\cite{Fadrny:2024}, reproduced in Fig.~\ref{fig:Fadrny2024_figure_5} here). (d) The relative difference $(\langle \cdot \rangle_{m=2} - \langle \cdot \rangle_{m=1}) / (\langle \cdot \rangle_{m=2} + \langle \cdot \rangle_{m=1})$ for $W_{1}$ (black circles), $D_{\rm KL}$ (blue asterisks), and $D_{\rm B}$ (red triangles) plotted as functions of $|\alpha|$. The plots corresponding to $\langle D_{\rm KL} \rangle$ and $\langle D_{\rm B} \rangle$ are essentially similar. Overall, $\langle W_{1} \rangle$, $\langle D_{\rm KL} \rangle$ and $\langle D_{\rm B} \rangle$ computed directly from the tomograms, mirror the trends in the fidelity computed after state reconstruction. Since the relative differences are more pronounced with increasing $|\alpha|$ for $D_{\rm KL}$ and $D_{\rm B}$ compared to $W_{1}$, they are better discriminators of photon addition to the CS for $|\alpha| > 1$.}
\label{fig:avg_WD_D_KL_div_D_BD_rel_diffs_CS_beta_opt_PACS_m_1_2_N_5_quads_vary_alpha}
\end{figure}
\noindent
(ii) {\it The fidelity of an $m$-PACS $|\alpha,m\rangle$ with a CS of optimal  amplitude $\beta_{\rm opt} = |\alpha| [1+(1+4m/|\alpha|^{2})^{1/2}]/2$, namely $|\langle \beta_{\rm opt} | \alpha,m \rangle|^{2}$, plotted as functions of $|\alpha|$ for $m=1$ and $2$}: 
A CS with this value of $\beta_{\rm opt}$ has maximum fidelity with $|\alpha,m\rangle$. These plots, obtained both theoretically and from the state reconstructed from the experimental data (Figs.~5(a) and~(b) respectively of~\cite{Fadrny:2024}, reproduced in Fig.~\ref{fig:Fadrny2024_figure_5} here), show that the fidelity increases with the value of $m$ for a given value of $|\alpha|$.
Further, it is a monotonically increasing function of $|\alpha|$, asymptotically approaching unity.

As an alternative to this procedure, we have computed $W_{1}$, $D_{\rm KL}$ and $D_{\rm B}$ between $|\alpha,m\rangle$ ($m=1,2$) and the CS $|\beta_{\rm opt}\rangle$ as functions of $|\alpha|$, in different tomographic slices (i.e., for different values of $\theta$ in the corresponding tomograms).

We have verified that the overall manner in which each of these quantities changes with $|\alpha|$ is similar in all the quadratures. Hence $\langle W_{1} \rangle$, $\langle D_{\rm KL} \rangle$ and $\langle D_{\rm B} \rangle$ are shown in Figs.~\ref{fig:avg_WD_D_KL_div_D_BD_rel_diffs_CS_beta_opt_PACS_m_1_2_N_5_quads_vary_alpha}(a),~(b) and~(c) here, where $\langle \cdot \rangle$ denotes an average over $5$ different tomographic slices, i.e., $5$ values of $\theta$ equally spaced between $0$ and $\pi$. We have verified that $5$ such values suffice. We observe that the numerical values of these three averages increase with increasing $m$, for any given $|\alpha|$, and also decrease with increase in $|\alpha|$. Our results are therefore in agreement with the trends seen in the fidelity plots. In order to compare the performance of the three quantifiers, we define the corresponding relative difference in each case as $(\langle \cdot \rangle_{m=2} - \langle \cdot \rangle_{m=1})/(\langle \cdot \rangle_{m=2} + \langle \cdot \rangle_{m=1})$. These relative differences are plotted in Fig.~\ref{fig:avg_WD_D_KL_div_D_BD_rel_diffs_CS_beta_opt_PACS_m_1_2_N_5_quads_vary_alpha}(d) for $W_{1}$ (black circles), $D_{\rm KL}$ (blue asterisks) and $D_{\rm B}$ (red triangles) as functions of $|\alpha|$. It is clear that the plots corresponding to $\langle D_{\rm KL} \rangle$ and $\langle D_{\rm B} \rangle$ are essentially similar. Thus, computing any one of the three quantities $\langle W_{1} \rangle$, $\langle D_{\rm KL} \rangle$ or $\langle D_{\rm B} \rangle$ directly from the tomograms, provides a viable method to assess the trends in the fidelity between $|\alpha,m\rangle$ ($m=1,2$) and $|\beta_{\rm opt}\rangle$ as functions of $|\alpha|$. As $|\alpha|$ increases, however, it is seen from Fig.~\ref{fig:avg_WD_D_KL_div_D_BD_rel_diffs_CS_beta_opt_PACS_m_1_2_N_5_quads_vary_alpha}(d) that $D_{\rm KL}$ and $D_{\rm B}$ turn out to be better discriminators between the addition of $1$ versus $2$ photons to the CS, than $W_{1}$. \\

\begin{figure}
\centering
\includegraphics[width=0.45\textwidth]{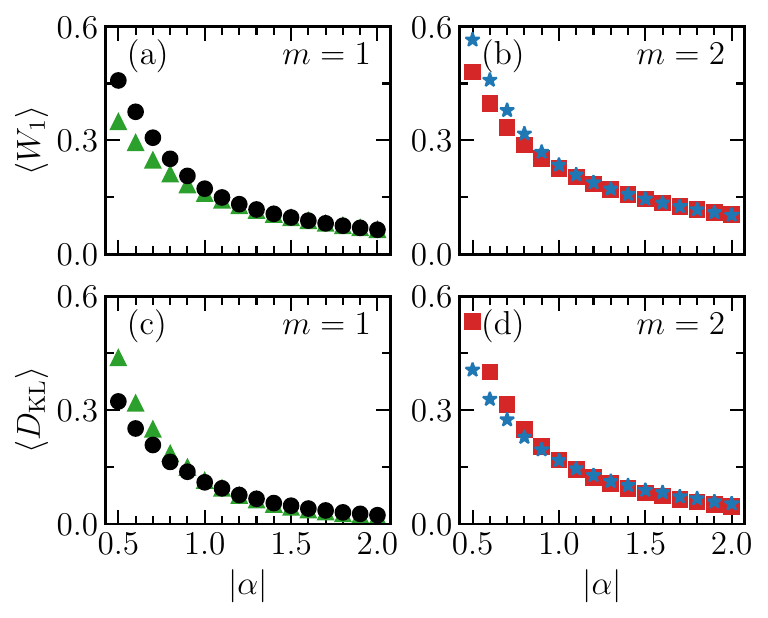}
\caption{Top panel: $\langle W_{1} \rangle$ between $|\alpha,m\rangle$ and $|{\sf g}_{m}(\alpha)\alpha\rangle$ ($m=1$: green triangles and $m=2$: red squares), and between $|\alpha,m\rangle$ and $|\beta_{\rm opt}\rangle$ ($m=1$: black circles and $m=2$: blue asterisks), as functions of $|\alpha|$. Bottom panel: $\langle D_{\rm KL} \rangle$ between $|\alpha,m\rangle$ and $|{\sf g}_{m}(\alpha)\alpha\rangle$ ($m=1$: green triangles and $m=2$: red squares), and between $|\alpha,m\rangle$ and $|\beta_{\rm opt}\rangle$ ($m=1$: black circles and $m=2$: blue asterisks), as functions of $|\alpha|$.}
\label{fig:avg_W1_D_KL_div_comparison_g_m_alpha_beta_opt_PACS_m_1_2}
\end{figure}
\noindent
(iii) {\it The fidelities between $|\alpha,m\rangle$ and a CS with amplitude ${\sf g}_{m}(\alpha) \alpha$ as functions of  $|\alpha|$,  shown in Figs.~5(a) and~(b) of~\cite{Fadrny:2024} (reproduced in Fig.~\ref{fig:Fadrny2024_figure_5} here)}: Here too, the reconstructed density matrices of the experimentally obtained state have been used to determine the fidelities. From these plots it it clear that for $m=1$ and $2$, both the theoretical computation using expressions for these two states, and the experimental data, show an increase in fidelity with increasing $|\alpha|$. Further, for a given value of $m$ and for small values of $|\alpha|$, the fidelity of $|\alpha,m\rangle$ with the CS $|\beta_{\rm opt}\rangle$ is larger than the corresponding fidelities with $|{\sf g}_{m}(\alpha) \alpha\rangle$. For values of $|\alpha| > 1$, these two fidelities are comparable and asymptotically reach unity in both cases. These features follow by noting that in the limit $|\alpha|\rightarrow 0$, $\beta_{\rm opt} \rightarrow \sqrt{m}$, whereas ${\sf g}_{m}(\alpha)\alpha \rightarrow 0$. For large $|\alpha|$, on the other hand, $\beta_{\rm opt} \approx {\sf g}_{m}(\alpha)\alpha$.
Noiseless amplification therefore works well for $|\alpha| \gtrsim 1$.

As an alternative, we have computed $W_{1}$, $D_{\rm KL}$ and $D_{\rm B}$ between $|\alpha,m\rangle$ and $|{\sf g}_{m}(\alpha)\alpha\rangle$ ($m=1,2$) as functions of $|\alpha|$, from their respective tomograms. See plots in Figs.~\ref{fig:avg_W1_D_KL_div_comparison_g_m_alpha_beta_opt_PACS_m_1_2}(a)--(d) for $W_{1}$ and $D_{\rm KL}$ with $m=1$ (green triangles) and $m=2$ (red squares). We have verified that $D_{\rm B}$ follows the same trend as $D_{\rm KL}$. In the same figures, for ready comparison we have also included the corresponding plots from Fig.~\ref{fig:avg_WD_D_KL_div_D_BD_rel_diffs_CS_beta_opt_PACS_m_1_2_N_5_quads_vary_alpha} between $|\alpha,m\rangle$ and $|\beta_{\rm opt}\rangle$ ($m=1$, black circles; $m=2$, blue asterisks). It is clear that $W_{1}$, $D_{\rm KL}$ and $D_{\rm B}$ reflect the experimental results stated in the preceding paragraph for $|\alpha| \gtrsim 1$. Hence, in the noiseless amplification regime all the three quantifiers efficiently capture the fidelity trends. However, for $|\alpha| \lesssim 1$, whereas the plots for $D_{\rm KL}$ are consistent with the experimental result (in the sense that for a given value of $m$, $D_{\rm KL}$ between $|\alpha,m\rangle$ and $|\beta_{\rm opt}\rangle$ is smaller than $D_{\rm KL}$ between $|\alpha,m\rangle$ and $|{\sf g}_{m}(\alpha)\alpha\rangle$), $W_{1}$ marginally departs from this feature. This is readily explained by noting that $W_{1}$ (the ``earth mover's distance") is the transportation cost of transforming one PDF to another. For $m=1$, in the case of $|\beta_{\rm opt}\rangle$ $W_{1}$ tends to the distance between $|1\rangle$ and the CS $|\alpha=1\rangle$, whereas in the case of $|g_{1}(\alpha)\alpha\rangle$ $W_{1}$ tends to the corresponding distance between $|1\rangle$ and $|0\rangle$. The transportation distance between a Gaussian and a shifted Gaussian is responsible for the departure mentioned above. A similar argument holds for higher values of $m$.

The plots in Fig.~\ref{fig:Fadrny2024_figure_5} shed light on important aspects of classical versus quantum amplification. As expected, the tomograms corresponding to $|{\sf g}_{m}(\alpha)\alpha\rangle$ and $|\alpha,m\rangle$ are visually similar for sufficiently large $|\alpha|$ (see Fig.~9 of SM). However, the difference between the amplified CS and the $m$-PACS are borne out by computing the corresponding variances in the $x$-quadrature (Fig.~6 of SM). The former is a minimum uncertainty state (an amplification gained `classical' state). In the latter case, while addition of a photon amplifies the state, it also leads to squeezing.

\section{Summary and outlook\label{sec:Summary}}
Addition of photons to coherent light is an area of immense current interest as it leads to nonclassical properties. We demonstrate that the number of photons added to the CS $|\alpha\rangle$ can be readily identified by inspection of appropriate tomograms, and, if necessary, from variances in conjugate quadratures computed from the tomograms (see SM). Further, higher moments of the quadrature variables, and the mean photon number can also be obtained solely from the tomograms, circumventing detailed state reconstruction. Apart from this, we have investigated the manner in which the distance between PDFs corresponding to different photon added states vary with relevant parameters. This is done using markers ($W_{1}$, $D_{\rm KL}$ and $D_{\rm B}$) which are readily computable from relevant tomograms.

In particular, we have considered one and two photon addition to the CS $|\alpha\rangle$, and compared our results with experimental findings. We have established that the manner in which all these markers vary with $\alpha$ are similar, and mirror recent experimental findings where fidelities have been computed from the reconstructed state. 

We have also examined the role of photon addition and subtraction to nonclassical light such as the single mode squeezed vacuum and cat states. In this case, we have shown that these markers differ in the manner in which they vary with relevant parameters, and we have indicated their individual advantages.
 
While for single mode systems state reconstruction is relatively easy, this is not true in general for multimode systems. Our work therefore opens up possibilities for extending this tomographic approach to examine such systems.

\begin{acknowledgments} 
We acknowledge partial support through funds from Mphasis to the Centre for Quantum Information, Communication and Computing (CQuICC), Indian Institute of Technology Madras. SL and VB thank the Department of Physics, Indian Institute of Technology Madras for infrastructural support.
\end{acknowledgments}

\bibliography{references}

\end{document}


\title{Supplementary Material for ``Tomographic markers and photon addition to coherent states of light: Comparison with experiment''}

\author{Soumyabrata Paul}
\email{soumyabrata@physics.iitm.ac.in}
\affiliation{Department of Physics, Indian Institute of Technology Madras, Chennai 600036, India}
\affiliation{Center for Quantum Information, Communication and Computing (CQuICC), Indian Institute of Technology Madras, Chennai 600036, India}
\author{S. Lakshmibala}
\affiliation{Center for Quantum Information, Communication and Computing (CQuICC), Indian Institute of Technology Madras, Chennai 600036, India}
\author{V. Balakrishnan}
\affiliation{Center for Quantum Information, Communication and Computing (CQuICC), Indian Institute of Technology Madras, Chennai 600036, India}
\author{S. Ramanan}
\affiliation{Department of Physics, Indian Institute of Technology Madras, Chennai 600036, India}
\affiliation{Center for Quantum Information, Communication and Computing (CQuICC), Indian Institute of Technology Madras, Chennai 600036, India}

\date{\today}

\begin{abstract}
\end{abstract}

\maketitle


\section{Fock states: Tomograms and  Wigner functions\label{sec:Fock_states_tomograms_Wigner_functions}} 
\begin{figure}[h]
\centering
\includegraphics[width=0.45\textwidth]{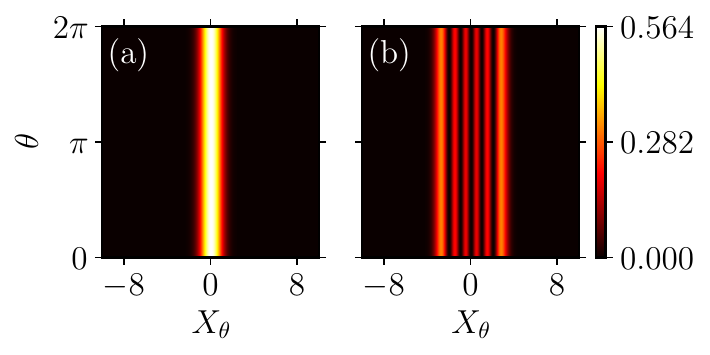}
\caption{Optical tomograms $w(X_{\theta}, \theta)$ corresponding to (a) the $0$-photon state and (b) the $5$-photon state. In general, for the $n$-photon state the tomogram has $n$ vertical dark cuts. Since the PDFs are independent of $\theta$, tomographic markers such as $W_{1}$, $D_{\rm KL}$ and $D_{\rm B}$ can be computed in any quadrature, without loss of generality.}
\label{fig:fig_Fock_tomogram_0_5}
\end{figure}

\begin{figure}[h]
\centering
\includegraphics[width=0.45\textwidth]{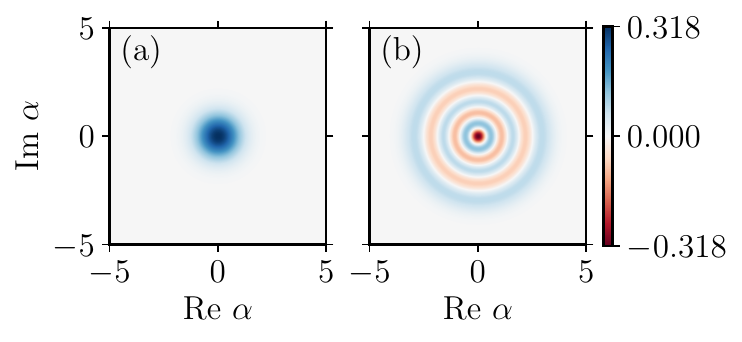}
\caption{Wigner functions corresponding to (a) the $0$-photon state and (b) the $5$-photon state. Here ${\rm Re}~\alpha = x/\sqrt{2}$ and ${\rm Im}~\alpha = p/\sqrt{2}$.}
\label{fig:fig_Fock_Wigner_0_5}
\end{figure}

Optical tomograms and their corresponding Wigner functions for two Fock states are shown in Figs.~\ref{fig:fig_Fock_tomogram_0_5} and~\ref{fig:fig_Fock_Wigner_0_5}  respectively.
We observe from Fig.~\ref{fig:fig_Fock_tomogram_0_5} that the $5$-photon state has $5$ vertical dark cuts, while the $0$-photon state has none. 
This is reflected in the Wigner functions through the appearance of five rings (in white) for the $5$-photon state, and the lack of such structure for the $0$-photon state.
For generic states, the PDFs and therefore the distance between PDFs computed from the tomograms will be $\theta$-dependent, although this is not so for photon number states, as seen from the tomographic patterns in Fig.~\ref{fig:fig_Fock_tomogram_0_5}.

\begin{figure}[h]
\centering
\includegraphics[width=0.28\textwidth]{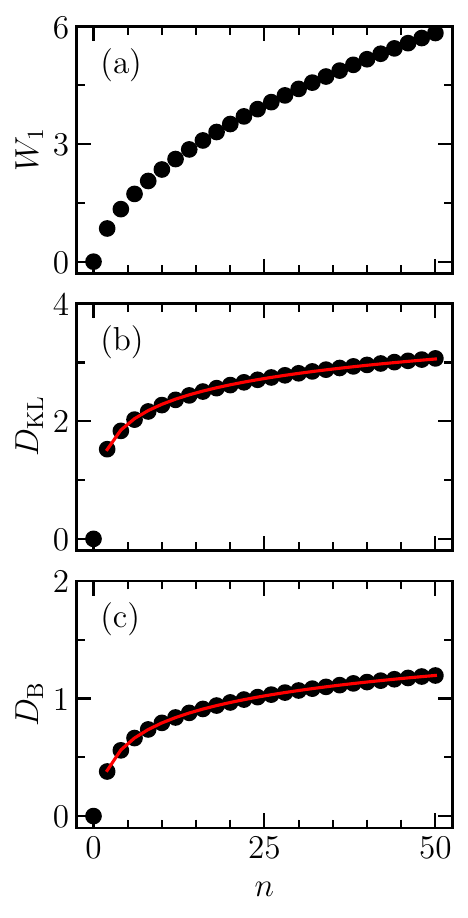}
\caption{Top to bottom: (a) $W_{1}(G_{0}, G_{n})$ versus the photon number $n$ computed along the $x$-quadrature for photon number states. $G_{0}$ and $G_{n}$ are respectively the CDFs corresponding to the $0$-photon state and the $n$-photon state. It can be seen that $W_1 \sim n^{1/2}$ for large $n$. (b) $D_{\rm KL}(g_{0}, g_{n})$ versus the photon number $n$ computed along the $x$-quadrature for photon number states (black circles). $g_{0}$ and $g_{n}$ are respectively the PDFs corresponding to the $0$-photon state and the $n$-photon state. The solid curve (red) corresponds to a fit with the function $0.478 \, \mathrm{ln} \,n + 1.187$. (c) $D_{\rm B}(g_{0}, g_{n})$ versus the photon numner $n$ computed along the $x$-quadrature for photon number states (black circles). $g_{0}$ and $g_{n}$ are respectively the PDFs corresponding to the $0$-photon state and the $n$-photon state. The solid curve (red) corresponds to a fit with the function $0.251 \,\mathrm{ln} \, n + 0.214$.}
\label{fig:Fock_WD_DKL_DB}
\end{figure}

\begin{figure}
\centering
\includegraphics[width=0.35\textwidth]{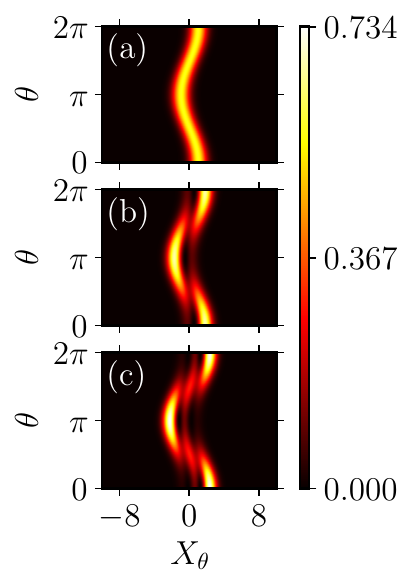}
\caption{Optical tomograms corresponding to (a) $|\alpha\rangle$, (b) $|\alpha,1\rangle$ and (c) $|\alpha,2\rangle$ for $\alpha=0.7$. Note that the tomogram for $|\alpha,1\rangle$ has one vertical cut, and that for $|\alpha,2\rangle$ has two vertical cuts. In general, it can be seen that the tomogram for $|\alpha,m\rangle$ has $m$ vertical cuts.}
\label{fig:fig_tomogram_CS_1_PACS_2_PACS_alpha_0.7}
\end{figure}

\section{Normal-ordered moments from optical tomograms\label{sec:Wunsche_procedure}}
We summarize the key steps in the procedure~\cite{Wunsche:1996} for obtaining the normal-ordered moments for any infinite-dimensional single-mode system from its optical tomogram.

Any single-mode density operator ${\hat \rho}$ can be written in the normal-ordered form
\begin{equation}
{\hat \rho} = \sum_{m=0}^{\infty}\sum_{n=0}^{\infty} {\hat \rho}_{m, n} {\hat a}^{\dagger m} {\hat a}^{n}.
\label{eq:eq_1}
\end{equation}
Here
\begin{equation}
{\hat \rho}_{m,n} = \sum_{k=0}^{\{m, n\}} \frac{(-1)^{k}}{k!\sqrt{(m-k)!(n-k)!}} \langle m-k|{\hat \rho}|n-k \rangle,
\label{eq:eq_2}
\end{equation}
and $\{m,n\}$ stands for ${\rm min}(m, n)$.

The single-mode density operator ${\hat \rho}$ can also be written as
\begin{equation}
{\hat \rho}=\sum_{m,n=0}^{\infty}|m\rangle \langle n| {\rm Tr}\left( |m\rangle \langle n| {\hat \rho} \right).
\label{eq:eq_3}
\end{equation}
Using the projector operator expansion
\begin{equation}
|m\rangle \langle n| = \frac{1}{\sqrt{m!n!}} \sum_{s=0}^{\infty} \frac{(-1)^{s}}{s!} {\hat a}^{\dagger m+s} {\hat a}^{n+s},
\label{eq:eq_4}
\end{equation}
in~\eqref{eq:eq_3} gives
\begin{equation}
{\hat \rho}=\sum_{m=0}^{\infty}\sum_{n=0}^{\infty} {\mathcal A}_{m,n} {\rm Tr} \left( {\hat a}^{\dagger m} {\hat a}^{n} {\hat \rho} \right),
\label{eq:eq_5}
\end{equation}
where
\begin{equation}
{\mathcal A}_{m,n}=\sum_{k=0}^{\{m,n\}} \frac{(-1)^{k}}{k!\sqrt{(m-k)!(n-k)!}}|m-k\rangle \langle n-k|.
\label{eq:eq_6}
\end{equation}
Again, using Eqs.~(\ref{eq:eq_5}),~(\ref{eq:eq_6}),
\begin{equation}
\langle X_\theta, \theta | m \rangle \langle n | X_\theta, \theta \rangle = \frac{e^{-X_\theta^2}}{\sqrt{\pi}} \frac{e^{-i(m-n)\theta}}{2^{(m+n)/2}\sqrt{m!n!}} H_m(X_\theta) H_n(X_\theta),
\label{eq:eq_7}
\end{equation}
and the following property of the Hermite polynomials
\begin{equation}
H_{m+n}(X_\theta) = \sum_{k=0}^{\{m,n\}} \frac{(-2)^k m! n! H_{m-k}(X_\theta) H_{n-k}(X_\theta)}{k! (m-k)! (n-k)!},
\label{eq:eq_8}
\end{equation}
we can show that
\begin{align}
w(X_\theta, \theta) &= \langle X_\theta, \theta | \rho | X_\theta, \theta \rangle \nonumber \\
&= \frac{e^{-X_\theta^2}}{\sqrt{\pi}} \sum_{m,n=0}^{\infty} \frac{e^{i(m-n)\theta}}{2^{(m+n)/2}\sqrt{m!n!}} H_{m+n}(X_\theta) \times \nonumber \\ 
& \quad\quad\quad\quad\quad\quad\quad \text{Tr}({\hat a}^{\dagger m} {\hat a}^{n} {\hat \rho}).
\label{eq:eq_9}
\end{align}
Finally, using the orthonormality property of the Hermite polynomials together with the expression
\begin{equation}
\sum_{k=0}^{n} \exp\left(2\pi iks/(n+1)\right) = (n+1)\delta_{s,0},
\label{eq:eq_10}
\end{equation}
in~\eqref{eq:eq_9} gives
\begin{align}
\langle {\hat a}^{\dagger m}{\hat a}^{n} \rangle &= {\rm Tr}({\hat a}^{\dagger m}{\hat a}^{n} \rho) \nonumber \\
&= C_{mn} \sum_{k=0}^{m+n} \exp \Big\{ - \frac{ik(m-n)\pi}{m+n+1} \Big\} \times\nonumber \\
& \int_{-\infty}^{\infty} dX_{\theta}~w \Big( X_{\theta}, \frac{k\pi}{m+n+1} \Big) H_{m+n}(X_{\theta}),
\label{eq:eq_11}
\end{align}
where
$C_{mn}=(m!n!)/\big\{ (m+n+1)!2^{(m+n)/2} \big\}$.

\section{Computation of the quadrature variance directly from the optical tomogram corresponding to any single-mode state of light and comparison with experimental plots\label{sec:variance_pacs_m_1_2}}
\begin{figure}
\centering
\includegraphics[width=0.40\textwidth]{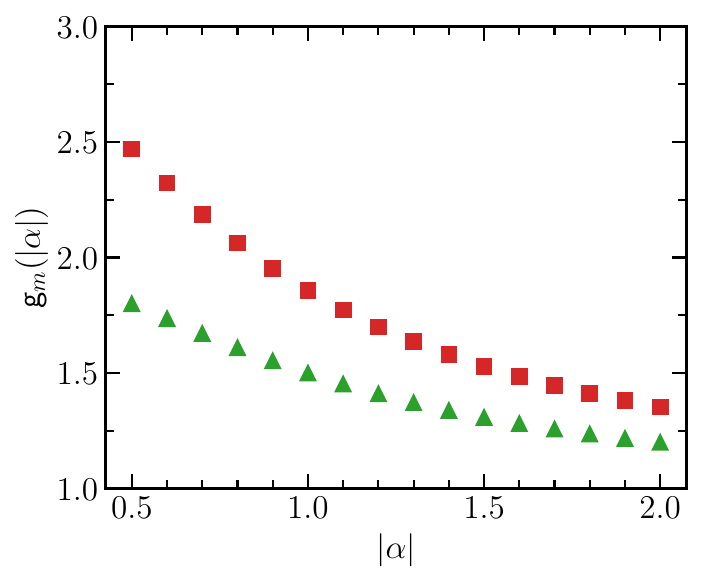}
\caption{The amplification gain ${\sf g}_{m}(|\alpha|)$ for $m=1$ (green triangles) and $m=2$ (red squares) versus $|\alpha|$, computed from the tomograms. These plots are in  good agreement with Figs.~4(a) and~(b) of~\cite{Fadrny:2024} computed from the reconstructed states, and reproduced in Figs.~\ref{fig:Fadrny2024_figure_4}(a) and~(b) here.}
\label{fig:g_pacs_m_1_2_vary_alpha}
\end{figure}
\begin{figure}
\centering
\includegraphics[width=0.45\textwidth]{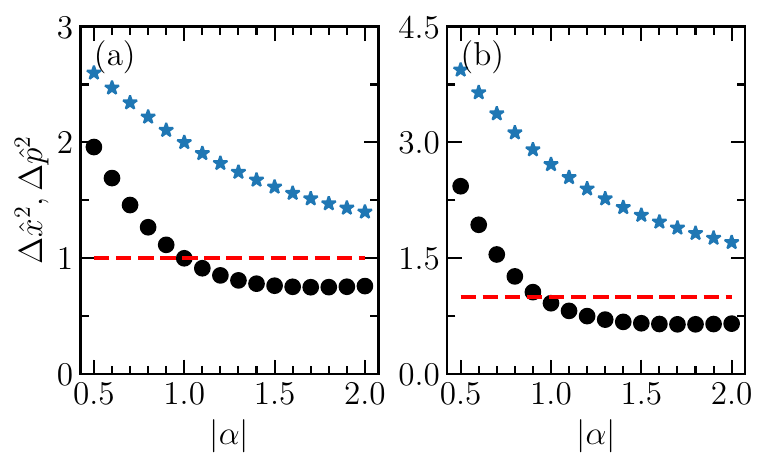}
\caption{The quadrature variances $\Delta{\hat x}^{2}$ (black circles) and $\Delta{\hat p}^{2}$ (blue asterisks) for the state (a) $|\alpha,1\rangle$, and (b) $|\alpha,2\rangle$ as functions of $|\alpha|$, computed {\it directly} from the optical tomograms. These plots agree well with the corresponding plots from the experiment ($0.5 \leqslant |\alpha| \leqslant 2$) reported in~\cite{Fadrny:2024}, shown in Figs.~\ref{fig:Fadrny2024_figure_4}(c) and~(d) respectively (purple dash-dot lines and blue dashed lines). The quadrature variance for the state $|\alpha\rangle$ is set to $1$ and is denoted by the red dashed line. For $|\alpha|=1.0$, $\Delta {\hat x}^{2}$ for both $|\alpha\rangle$ and $|\alpha,1\rangle$ are equal (see (a)). For $|\alpha| > 1$, $\Delta {\hat x}^{2} < 1$, denoting that both the states $|\alpha,1\rangle$ and $|\alpha,2\rangle$ are squeezed along the $x$-quadrature (see (a) and (b)). It is clear that there is no squeezing along the $p$-quadrature for the range of $|\alpha|$ considered.}
\label{fig:moments_pacs_m_1_2_vary_alpha}
\end{figure}
From Secs.~1 and~2 of the main text, we see that any tomogram is a collection of bonafide one-dimensional probability distribution functions (PDFs). Each PDF corresponds to a different value of $\theta$, the quadrature angle. For instance, $\theta=0$ corresponds to the $x$-quadrature, and $|\psi(x)|^{2}$ can be read off directly from the tomogram. Any expectation value of the form $\langle{\hat x}^{k}\rangle$ ($k>0$) can be readily computed now, as it involves evaluating $\int_{-\infty}^{\infty} dx~|\psi(x)|^{2} x^{k}$. The variance and the higher moments can therefore be directly obtained. Alternatively, one can also use the method outlined in Sec.~\ref{sec:Wunsche_procedure}. A similar exercise can be carried out for computing the variance and higher moments corresponding to any other quadrature.

\begin{figure}
\centering
\includegraphics[width=0.37\textwidth]{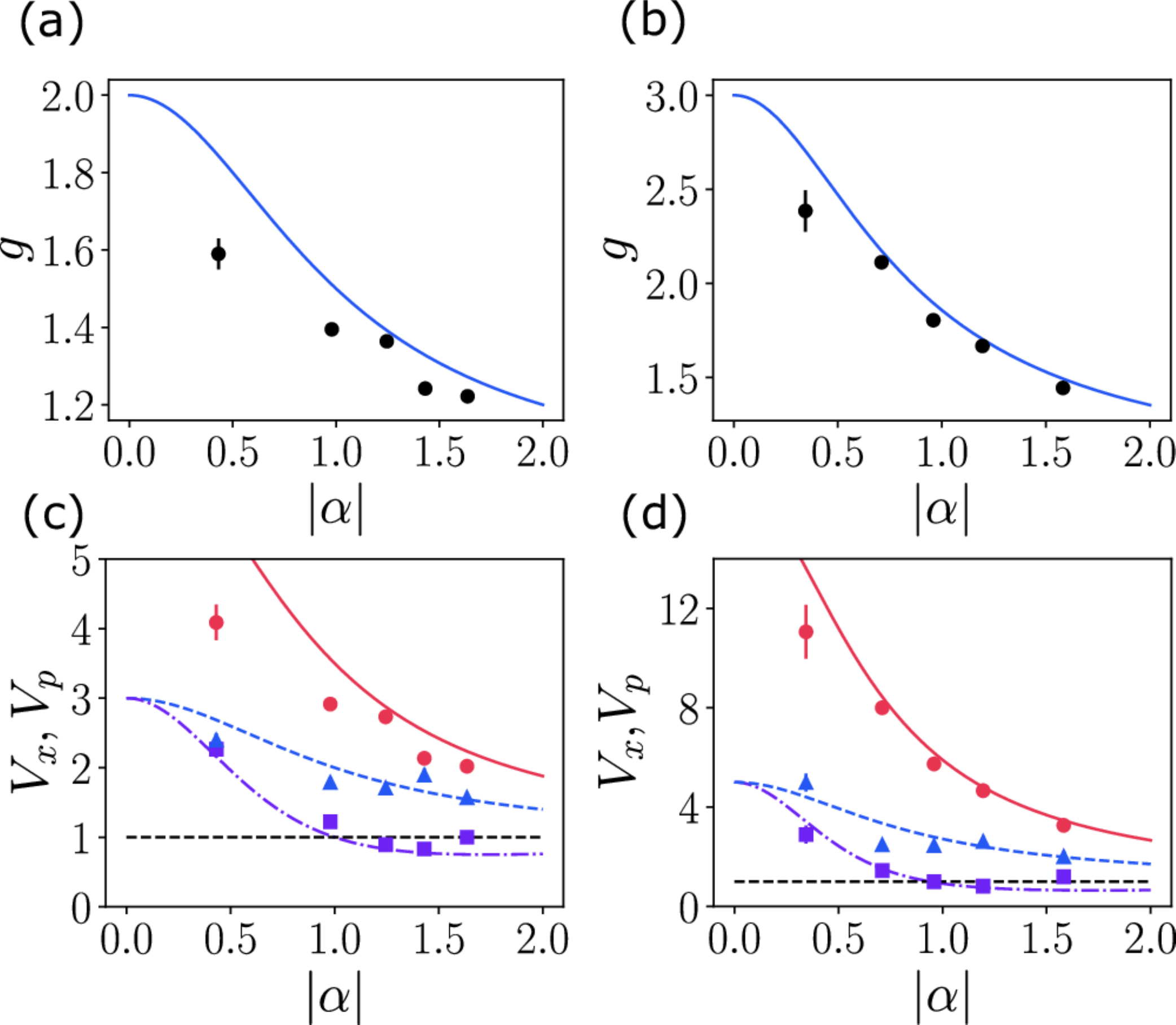}
\caption{{\bf Characterization of noiseless amplification of coherent states by $n$-photon addition}. {\bf a, b} The theoretical amplification gains ${\sf g}_{1}(\alpha)$ and ${\sf g}_{2}(\alpha)$ are plotted as lines and dots mark the experimental data. {\bf c, d} Experimentally determined variances of amplitude quadratures $V_{x}$ (purple squares) and phase quadratures $V_p$ (blue triangles) of the amplified states are plotted together with the theoretical dependencies (purple dot-dashed lines and blue dashed lines). As a benchmark, the red solid lines in (c, d) indicate quadrature variance achievable by deterministic amplifier with the theoretical gains ${\sf g}_{1}(\alpha)$ and ${\sf g}_{2}(\alpha)$, respectively. The red dots show variances for deterministic amplification with the experimentally observed gains. Vacuum variance is set to $1$ and indicated by the black dashed line. The error bars represent one standard deviation. For most of the data, the error bars are smaller than the symbol size. Data are plotted for $n = 1$ ({\bf a, c}) and $n = 2$ ({\bf b, d}). \\
\vspace{1ex}
This figure (with caption) has been taken from the publication ``Experimental preparation of multiphoton-added coherent states of light" by J. Fadrn{\'y}, M. Neset, M. Bielak, M. Je{\v{z}}ek, J. B{\'i}lek, and J. Fiur{\'a}{\v{s}}ek, npj Quantum Information {\bf 10}, 89 (2024)~\cite{Fadrny:2024}, DOI:10.1038/s41534-024-00885-y.}
\label{fig:Fadrny2024_figure_4}
\end{figure}

For our purpose we have computed the variances in both the $x$ and the $p$ (i.e., $\theta=\pi/2$) quadratures as functions of $|\alpha|$, directly from the tomograms corresponding to the $1$-PACS $|\alpha, 1\rangle$ and the $2$-PACS $|\alpha, 2\rangle$. Representative tomograms for these two states with $\alpha=0.7$ are shown in Figs.~\ref{fig:fig_tomogram_CS_1_PACS_2_PACS_alpha_0.7}(b) and~(c). These variance plots are shown in Fig.~\ref{fig:moments_pacs_m_1_2_vary_alpha}. They agree well with the corresponding plots obtained from reconstructed states in a recent experiment~\cite{Fadrny:2024} shown in Fig.~\ref{fig:Fadrny2024_figure_4}.

\section{Distance between photon number states from the tomogram\label{sec:Fock_states}}
We now calculate $W_1$, $D_{\rm KL}$ and $D_{\rm B}$ between the vacuum and the $n$-photon state. In general, the PDFs and therefore the distances computed from the tomograms
will be $\theta$-dependent, although this is not so in this case, as can be seen from the tomographic pattern in Sec.~\ref{sec:Fock_states_tomograms_Wigner_functions} here. We therefore compute $W_1$, $D_{\rm KL}$ and $D_{\rm B}$ in the $x$ quadrature, without loss of generality. The normalized PDF in this basis is given by the modulus squared of the oscillator wave function, namely $g_{n}(x) =  e^{-x^{2}} H_{n}^2(x)/ [2^{n} \pi^{1/2} n!]$, where $H_{n}(x)$ is the Hermite polynomial of order $n$, with a corresponding CDF $G_{n}(x)$. We have computed $W_{1}(G_{0}, G_{n})$, $D_{\rm KL}(g_{0}, g_{n})$ and $D_{\rm B}(g_{0}, g_{n})$, as functions of $n$, using Eqs.~(2)--(4) of the main text. The leading large $n$ behavior of $W_{1}$ (see Fig.~\ref{fig:Fock_WD_DKL_DB}(a)) is $\sim {n}^{1/2}$. This is to be expected on physical grounds: most of the probability mass is concentrated around $x = 0$ in the vacuum state, and around $\pm n^{1/2}$ in the $n$-photon state, for both even and odd $n$. This feature becomes more pronounced with increasing $n$. 
Further, we get
\begin{equation}
D_{\rm KL}(g_{0}, g_{n}) = \mathrm{ln} \,(2^{n} n!) - \frac{2}{\sqrt{\pi}} \int_0^\infty \!\!\! dx\, e^{-x^2} \, \mathrm{ln}\, H_{n}^2(x),
\end{equation}
\begin{equation}
D_{\rm B}(g_{0}, g_{n}) = -\mathrm{ln} \, \Big\{\frac{2}{\sqrt{2^{n} \, n! \, \pi}}\, \int_0^\infty \!\!\! dx \, e^{-x^2} \, |H_{n}(x)| \Big\}.
\end{equation}
\noindent Figures~\ref{fig:Fock_WD_DKL_DB}(b) and~\ref{fig:Fock_WD_DKL_DB}(c) show the variation of $D_{\rm KL}$  and $D_{\rm B}$ with $n$.
Using the leading behavior of $H_{n}(x)$ and Stirling's formula, it can be shown that the large-$n$ behavior of both these quantities $\sim  \ln n$. This logarithmic dependence renders them less sensitive {\it asymptotically} to the actual states that are involved in the comparison, in marked contrast to the $n^{1/2}$ {\it asymptotic} behavior of $W_1$. However, for small values of $n$ all the three quantifiers capture the trends effectively.

\section{Addition of photons to nonclassical light\label{sec:photons_added_nonclassical_light}} 
The $m$-photon added squeezed vacuum state $|\xi,m\rangle$ can be expanded in the Fock basis as
\begin{equation}
|\xi,m\rangle = \frac{1}{{\mathcal{N}^{1/2}_{m}}} \sum_{n=0}^{\infty} \frac{\sqrt{(2n + m)}!}{2^n n!} e^{i n \phi} (- \tanh r)^{n}~ |2n+m\rangle,
\label{eq:xi_m_photon_added_svs}
\end{equation}
with $\mathcal{N}_{m} = m! (\cosh r)^{m+1} P_{m}(\cosh r)$, where $P_{m}$ is the Legendre polynomial of degree $m$ ($=0,1,2, \dots$). Here $\xi=r\exp(i\phi)$, $r \geqslant 0$ and $\phi \in [0, 2\pi)$. $m=0$ corresponds to the squeezed vacuum state (SVS) $|\xi\rangle$.

In the case of the SVS, there are differences in the performance of the three tomographic markers $W_{1}$, $D_{\rm KL}$ and $D_{\rm B}$. We can see that while the number of added photons ($m=1$ or $m=2$), can be identified visually from the intensity shifts in the neighborhood of the $x$-quadrature of the corresponding tomograms, $W_{1}$ is substantially sensitive to the magnitude of the squeezing parameter even when a single photon is added to the SVS, although crossover regions are present in this case (see Ref.~\cite{Soumyabrata:2025}).

\begin{figure}
\centering
\includegraphics[width=0.40\textwidth]{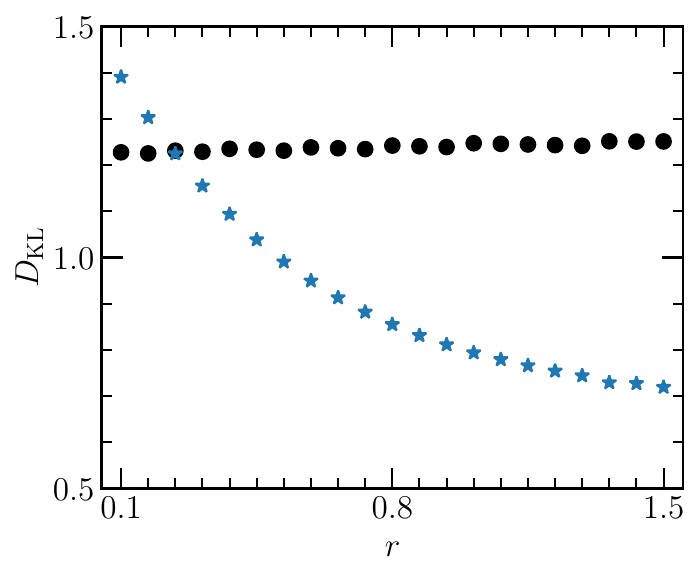}
\caption{$D_{\rm KL}$ between (a) the SVS $|\xi\rangle$ and the one photon added SVS $|\xi,1\rangle$ (black circles), and (b) $|\xi\rangle$ and the two photon added SVS $|\xi,2\rangle$ (blue asterisks) as a function of the squeezing parameter $r$. The squeezing is along the $x$-quadrature. The distances are calculated by comparing the PDFs along the $x$-quadrature ($\theta=0$) directly from the appropriate tomograms. Low sensitivity of $D_{\rm KL}$ to $r$ therefore signals addition of a single photon to the SVS, away from the crossover point $r \approx 0.24$ and its neighborhood.}
\label{fig:D_KL_SVS_SPASVS_DPASVS}
\end{figure}

However, $D_{\rm KL}$ is not sensitive to the squeezing parameter if one photon alone is added to the SVS (see Fig.~\ref{fig:D_KL_SVS_SPASVS_DPASVS}). This feature can be understood readily by noting that the logarithmic function in $D_{\rm KL}$ is not sufficiently sensitive to the squeezing parameter $r$. On addition of two photons, however, its sensitiveness to the magnitude of the squeezing parameter is considerably higher. The corresponding PDFs, and the location of the nodes in the PDFs are directly responsible for this behavior. This does not imply that $D_{\rm KL}$ is a bad marker, as it signals addition of a single photon if it does not change with the extent of squeezing, provided the crossover point at $r \approx 0.24$ and its neighborhood are avoided.

We have verified that for both one and two photon addition to the SVS, $D_{\rm B}$ follows the same trends as $D_{\rm KL}$.

A similar exercise has been carried out by us for the even coherent state $|\alpha\rangle_{+} = \big(|\alpha\rangle + |-\alpha\rangle\big)/\{2(1+\exp(-2|\alpha|^{2}))\}^{1/2}$ (where $|\alpha\rangle$ is a coherent state, and $\alpha \in \mathbb{C}$). In this case the number of added photons can be identified by inspecting the intensity shifts in the $p$~quadrature. We have also examined the manner in which the three markers vary with $|\alpha|$. We do not report the details here, as the performance of the three markers are similar to the case of photon addition to the CS.

\section{Classical and quantum amplification from tomograms\label{sec:classical_quantum_amplification}}

\begin{figure}
\centering
\includegraphics[width=0.40\textwidth]{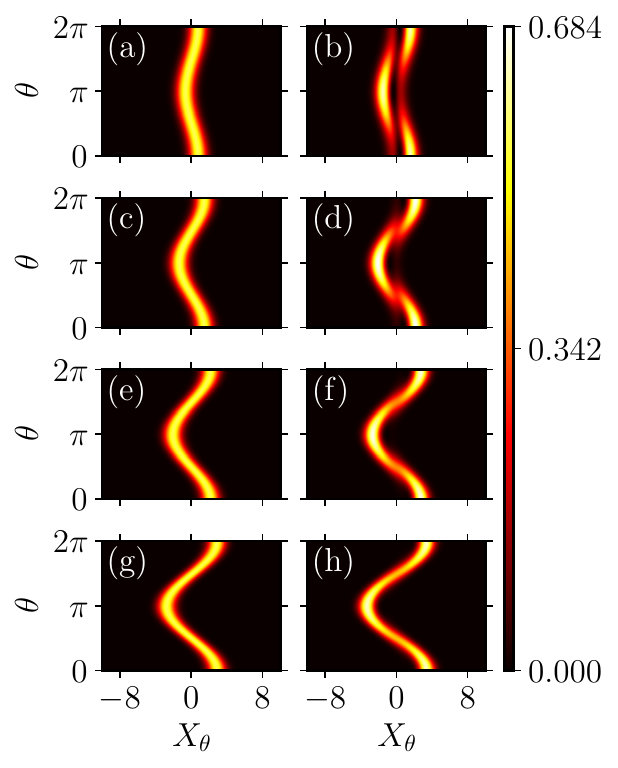}
\caption{Single-mode optical tomograms. Left column panels correspond to the coherent state $|{\sf g}_{1}(\alpha)\alpha\rangle$, and right column panels correspond to $|\alpha,1\rangle$. With $\alpha=0.5$ for (a) and (b), $1.0$ for (c) and (d), $1.5$ for (e) and (f), and $2.0$ for (g) and (h) respectively. Note that for $|\alpha| > 1$, the vertical cut in the tomograms corresponding to $|\alpha,1\rangle$ is absent, making them visually appear like the tomogram for the coherent state [see (e) -- (h)]. Hence for $|\alpha| > 1$ the fidelity between $|{\sf g}_{1}(\alpha)\alpha\rangle$ and $|\alpha,1\rangle$ is maximized (see Fig.~1 in the main text, which corresponds to the experiment reported in ``Experimental preparation of multiphoton-added coherent states of light" J. Fadrn{\'y} {\it et al.} npj Quantum Inf. {\bf 10}, 89 (2024)). However, the variances say, along the $x$-quadrature computed directly from these tomograms are different for the coherent state $|{\sf g}_{1}(\alpha)\alpha\rangle$ and for $|\alpha,1\rangle$ (see Fig.~\ref{fig:moments_pacs_m_1_2_vary_alpha} above).}
\label{fig:tomogram_PACS_m_1_g1_alpha_panel}
\end{figure}
It is common practice to identify a quantum state through its $P$ representation, Wigner function, etc. The Glauber-Sudarshan $P$ function associated with the state $|\alpha,m\rangle$ is given by~\cite{Agarwal:1991}
\begin{equation}
P(z) = \frac{\exp(|z|^2-|\alpha|^2)}{m! L_m(-|\alpha|^2)} \frac{\partial^{2m}}{\partial z^{*m} \partial z^m} \delta^{(2)}(z-\alpha),
\label{eq:pacs_p_function}
\end{equation}
where $L_{m}$ is the Laguerre polynomial of order $m$. The Wigner function is the inverse Radon transformation of the tomogram. These are generally computed after state reconstruction and can be used to distinguish between classical and quantum amplification. We outline a procedure to bypass state reconstruction and identify classical/quantum amplification from the tomogram itself. (See Fig.~\ref{fig:tomogram_PACS_m_1_g1_alpha_panel} and its caption, where tomograms for $|{\sf g}_{1}(\alpha)\alpha\rangle$ and $|\alpha,1\rangle$ are compared.) The single vertical cut in its tomogram is a signature of the one photon added CS $|\alpha,1\rangle$. However this disappears for higher value of $|\alpha|$. In particular, for $|\alpha|=2.0$ (Figs.~\ref{fig:tomogram_PACS_m_1_g1_alpha_panel}(g) and~(h)) the tomograms for the amplified CS $|{\sf }g_{1}(\alpha)\alpha\rangle$ and the quantum state $|\alpha,1\rangle$ are visually similar. However, their differences can be seen by computing the variances in the $x$-quadrature, directly from the tomograms as explained in Sec.~\ref{sec:Wunsche_procedure} here. Corresponding to the state $|{\sf }g_{1}(\alpha)\alpha\rangle$ the variance is minimum and independent of $|\alpha|$. In the case of the quantum state $|\alpha,1\rangle$ it can be seen from Fig.~\ref{fig:moments_pacs_m_1_2_vary_alpha}(a) that the variance is a function of $|\alpha|$, and further there is squeezing for $|\alpha| > 1$ (the reference level for squeezing is set to $1$). It turns out that for $|\alpha| = 1$, $\Delta {\hat x}^{2}$ equals unity both for the CS and $|\alpha,1\rangle$. Computation of $\Delta {\hat p}^{2}$ from the tomogram reveals that one is a CS although with amplification gain ${\sf g}_{1}(\alpha)$ whereas the other state is not a minimum uncertainty state. This exercise can be carried out for other values of $m$ as well.

\bibliography{references_SM}